\documentclass[11pt]{article}
\usepackage{amsfonts}

\usepackage{amssymb}
\usepackage{latexsym}
\usepackage{graphicx}
\usepackage[english]{babel}

\begin{document}
\input epsf

\makeatletter
\@addtoreset{equation}{section}
\makeatother


 \begin{center}
{\LARGE The information paradox and the infall problem}
\\
\vspace{18mm}
{\bf    Samir D. Mathur
\\}
\vspace{8mm}

\vspace{8mm}

Department of Physics,\\ The Ohio State University,\\ Columbus,
OH 43210, USA\\ 
mathur@mps.ohio-state.edu
\vspace{10mm}

\end{center}

\thispagestyle{empty}

\def\p{\partial}
\def\h{{1\over 2}}
\def\be{\begin{equation}}
\def\bea{\begin{eqnarray}}
\def\ee{\end{equation}}
\def\eea{\end{eqnarray}}
\def\d{\partial}
\def\la{\lambda}
\def\eps{\epsilon}
\def\bb{\bigskip}
\def\mm{\medskip}
\newcommand{\dm}{\begin{displaymath}}
\newcommand{\edm}{\end{displaymath}}
\renewcommand{\b}{\tilde{B}}
\newcommand{\gm}{\Gamma}
\newcommand{\ac}[2]{\ensuremath{\{ #1, #2 \}}}
\renewcommand{\ell}{l}
\newcommand{\z}{\ell}
\newcommand{\newsection}[1]{\section{#1} \setcounter{equation}{0}}
\def\bb{$\bullet$}
\def\Qbar{{\bar Q}_1}
\def\QPbar{{\bar Q}_p}
\def\q{\quad}
\def\bn{B_\circ}
\def\sq{{1\over \sqrt{2}}}
\def\z{|0\rangle}
\def\o{|1\rangle}
\def\sqi{{1\over \sqrt{2}}}

\let\a=\alpha \let\b=\beta \let\g=\gamma \let\d=\delta \let\e=\epsilon
\let\c=\chi \let\th=\theta  \let\k=\kappa
\let\l=\lambda \let\m=\mu \let\n=\nu \let\x=\xi \let\r=\rho
\let\s=\sigma \let\t=\tau
\let\vp=\varphi \let\vep=\varepsilon
\let\w=\omega      \let\G=\Gamma \let\D=\Delta \let\Th=\Theta
                     \let\P=\Pi \let\S=\Sigma

\def\h{{1\over 2}}
\def\t{\tilde}
\def\r{\rightarrow}
\def\nn{\nonumber\\}
\let\bm=\bibitem
\def\Kt{{\tilde K}}
\def\b{\bigskip}

\let\p=\partial
\def\u{\uparrow}
\def\d{\downarrow}

\begin{abstract}

\b

It is sometimes believed that small quantum gravity corrections to the Hawking radiation process can encode the correlations required to solve the black hole information paradox. Recently an inequality on the entanglement entropy of radiation was derived, which showed that such is not the case; one needs {\it order unity} corrections to low energy modes at the horizon to resolve the problem. In this paper we illustrate this inequality by a simple model where the state of the created Hawking pair at each stage is slightly modified by the state of the previous pair. The model can be mapped onto the 1-dimensional Ising chain and solved explicitly. In agreement with the general inequality we find that very little of the entanglement is removed by the encoded correlations. We then use the general inequality to argue that the black hole puzzles split into two problems: the `information paradox' and the `infall problem'. The former addresses the detailed state of low energy modes at the horizon and asks if these can be modified by order unity, while the latter asks for a coarse grained effective description of the infall of heavy observers into the degrees of freedom of the hole.

\end{abstract}
\vskip 1.0 true in

\newpage
\renewcommand{\theequation}{\arabic{section}.\arabic{equation}}

\def\p{\partial}
\def\r{\rightarrow}
\def\h{{1\over 2}}
\def\b{\bigskip}

\def\nn{\nonumber\\ }

\section{Introduction}
\label{intr}\setcounter{equation}{0}

Gedanken experiments tell us that black holes must have an entropy given in terms of their surface area \cite{bek}. This entropy suggests the existence of general thermodynamic behavior, and Hawking's computation of black hole radiation accords with this expectation, with the entropy, energy and temperature of the hole satisfying the first law of thermodynamics \cite{hawking1}.

\subsection{The entanglement problem}

But Hawking's computation of emission from the black hole leads to a serious difficulty. The emitted quanta are in fact members of particle pairs produced from the vacuum region near the horizon, with one member of the pair falling in to reduce the energy of the hole and the other member escaping as radiation to infinity. The members of this pair are in an entangled state, so the entanglement between the interior of the hole and infinity grows as the evaporation proceeds. When the black hole is reduced to a planck sized object the computation of Hawking cannot be trusted further, but the essential problem has already been created. If the evaporation proceeds to completion and the hole vanishes, then the quanta at infinity are left in a situation where they are in an entangled state, but there is nothing in the spacetime that they are entangled with. This fact lead to Hawking's suggestion that the general state of matter in a theory of gravity must be described, even at a fundamental level, by density matrices instead of pure state wavefunctions \cite{hawking2}. If, on the other hand, the evaporation terminates with a planck scale remnant, then we are led to the existence of an infinite number of possible remnant states, all within a bounded energy interval of order planck mass, and while this is not an impossible situation to imagine, it does create an awkward behavior for the quantum gravity theory.

\subsection{Black hole `hair'}

These difficulties would be avoided if we could somehow show that the process of radiation gets modified in a way such that the entanglement between the inside and outside of the hole decreases to zero as the hole becomes small. Hawking's computation assumed that the state in the vicinity of the horizon was the vacuum in the coordinates which are good at the horizon. If we could construct black holes with {\it other} states at the horizon then the Hawking computation would need modification, and 
entanglement may not monotonically increase. In fact when we burn a piece of paper the entanglement between the paper and the emitted radiation at first goes up, but after about half the paper has disappeared, this entanglement starts to go down, reaching zero when the paper finally disappears \cite{page}. In this process it is crucial that the radiated quanta `see' the detailed state of the paper in the process of being emitted, so that when the evaporation is completed, the information of the state of the paper is transferred to correlations among the emitted photons. By contrast, in Hawking's computation the emitted quanta emerged from the vacuum in a predetermined entangled state; this state is independent of the information  the matter that made the hole, and this is what forces the entanglement to continue rising all the way to the endpoint of evaporation.

But attempts to change the state at the horizon ran into trouble. People could not find deformations of the black hole: a small perturbation added to the black hole either falls into the singularity or escapes to infinity. Forcing the perturbation to be time independent leads to a divergent stress tensor at the horizon. This lead to the `no hair' conjecture: the only state possible for the hole is the standard vacuum geometry determined by the conserved charges \cite{nohair}. If we indeed cannot find
any hair, then we have the vacuum state at the horizon and would have to follow along with Hawking's computation and its startling conclusion.

\subsection{The question of small corrections}

At this point we can consider the following possibility. Hawking performed a leading order computation of radiation, using the semiclassical metric around the black hole horizon. This would appear to be a good approximation because the curvature at the horizon is low. But there will always be small subleading corrections from quantum gravity effects in the full quantum theory. We require that our quantum gravity theory have a good low curvature limit (`laboratory physics limit'). Thus these corrections will have to vanish when ${M\over m_{pl}}\r \infty$, where $M$ is the mass of the hole and $m_{pl}$ is the planck mass. But a large hole also emits a very large number of quanta; this number is $N\sim ({M\over m_{pl}})^2$ for a 3+1 dimensional Schwarzschild hole. We can then ask if it possible that the largeness of the number of quanta can offset the smallness of the corrections. In other words, while each Hawking pair would be emitted in almost the same state found by Hawking, the small corrections to this entangled pair state might generate a  cumulative effect which leads to an overall zero entanglement between the total set of radiated quanta and the total set of their partners inside the hole.

But in a recent paper \cite{mathurfuzz} it was shown that this line of thought does not work. It is  true of course that small quantum gravity corrections can come from various sources. We allow an arbitrary correction of this type.  Thus the state of each  newly created  Hawking pair  is given a small correction which can depend in any way on the state of the matter which made the hole and the state of all Hawking pairs that have been created at earlier steps. All one requires is that this correction be small, with a departure from the leading order Hawking pair state of order $\epsilon$. One then computes the entanglement between the set of radiated quanta and the state inside the hole after $N$ steps of emission. One finds that the entanglement does {\it not} reduce in any significant way, even though we take $N$  large. In the way the computation was set up in \cite{mathurfuzz}, the leading order Hawking computation led to an increase of entanglement by $\Delta S=\ln 2$ with the emission of each new Hawking pair. When we include the corrections, one finds that the entanglement still increases with each emission, with 
\be
\Delta S> \ln 2-2\epsilon
\label{one}
\ee
For a large hole (${M\over _{pl}}\gg 1$) we must have have $\epsilon\ll 1$ if there is indeed `no hair' at the horizon, and we see that the essence of Hawking's argument is unchanged; the entanglement will continue to rise until the hole becomes order planck size and the corrections need no longer be taken small.

\subsection{The computations of this paper}

The inequality derived in \cite{mathurfuzz} required the use of a standard result in quantum information theory: the strong subadditivity of entanglement entropy \cite{lieb}. While this result is well known, it does not have a simple derivation. Thus its consequences may not be always built into out intuitive thinking about quantum states. So in this paper we take a simple example to illustrate the inequality. We take a simple model for the radiation process, where the state of the Hawking pair at any time step is modified in a small way by the emission at the {\it previous} time step. Such a situation can be imagined for various models of black hole evaporation that have been considered. For example,  in \cite{pw} it was noted that if a quantum is emitted at one time step then the temperature of the hole is increased a little for the emission at the next step, and so the emission amplitudes at this latter step would be affected somewhat, generating small correlations among the emitted quanta. 

For the simple model that we set up, we are able to solve for the entanglement entropy after $N$ steps of evolution. This is achieved by mapping the problem to that of a 1-dimensional Ising chain, which can of course be solved exactly. We find that the inequality (\ref{one}) is satisfied, so that such correlations cannot resolve the Hawking problem.

We then discuss how the inequality (\ref{one}) leads to a separation of the problem into {\it two} problems.  For the low energy ($E\sim kT$) quanta emitted by the hole we {\it must} find hair at the horizon that modify the evolution of low energy modes by order {\it unity}, if we are to have any hope of removing the entanglement that leads to Hawking's problem. On the other hand for massive infalling objects (observers with $E\gg kT$) we can hope for a coarse grained physics that is insensitive to the details of the hair, and that allows an effective description that mimics infall into the Schwarzschild geometry. We end with a discussion of how in string theory one finds that black hole microstates are given by  states where the state at the `horizon' is {\it not} the vacuum, and thus the Hawking argument fails for the evolution of low energy modes. Collective excitations of these microstates can be described by `effective' degrees of freedom which might mimic free infall, but this `infall problem' is different from the issue of resolving the information paradox.

\section{The inequality}
\label{ineq}\setcounter{equation}{0}

In this section we review the nature of Hawking emission, and explain the setup used to derive the inequality (\ref{one}). For more details on this discussion, see \cite{mathurfuzz}.

\subsection{The Hawking process and entanglement entropy at the leading order}

\begin{figure}[htbp]
\begin{center}
\includegraphics[scale=.18]{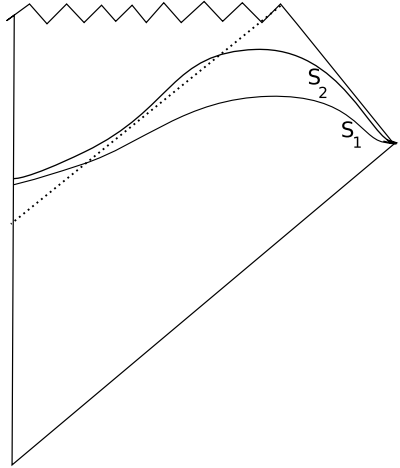}
\caption{{The Penrose diagram of a black hole formed by collapse of the `infalling matter'. The spacelike slices satisfy all the niceness conditions required for semiclassical evolution in a gravity theory.}}
\label{fthree}
\end{center}
\end{figure}

\begin{figure}[htbp]
\begin{center}
\includegraphics[scale=.18]{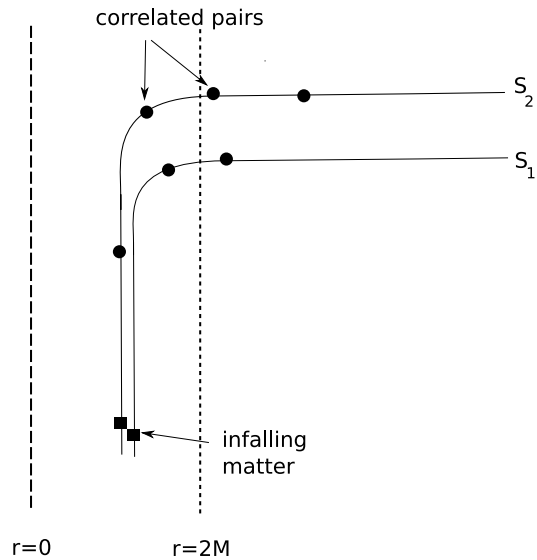}
\caption{{A schematic set of coordinates for the Schwarzschild hole. Spacelike slices are $t=const$ outside the horizon and $r=const$ inside. Infalling matter is very far from the place where pairs are created.}}
\label{ftwo}
\end{center}
\end{figure}

Consider a Schwarzschild hole with metric
\be
ds^2=-(1-{2M\over r})dt^2+{dr^2\over 1-{2M\over r}}+r^2 d\Omega_2^2
\label{qthree}
\ee
There is no singularity at the horizon, and the full spacetime made by a collapsing shell is given by the Penrose diagram shown in fig.\ref{fthree}. 
The Hawking process can be examined on the smooth slices shown in the figure, which capture the evolution up to any point where the horizon is still macroscopic,  without approaching the singularity anywhere. Since the conformal scaling of the Penrose diagram distorts the geometry of the slices, it is better to look at the slices shown in the schematic figure fig.\ref{ftwo}. The radial coordinate $r$ is plotted on the horizontal axis, and the other axis is just a way to depict evolution; it could be an Eddington-Finkelstein coordinate for instance.

Outside the hole we can make a spacelike slice by taking a $t=t_0$ surface. Inside the hole, we take a $r=r_0$ surface. These two parts of the surface can be joined by a smooth connection region ${\cal C}$.
To get evolution we must now construct a `later' slice. The $t=t_0$ part is simply changed to $t=t_0+\delta t$. Inside the hole, the forward evolution takes us to $r=r_0+\delta r$. Since we do not wish to get close to the singularity at $r=0$, we take $\delta r$ very small, such that in the entire evolution we are interested in we never get to, say, the region $r<{M\over 2}$. There is of course nothing wrong if we evolve one part of a spacelike slice forwards in time while not evolving another part of the slice; this is just the `many fingered time' of general relativity.

We now see that the $r=const$ part of this later slice has to be longer -- it has to `stretch' -- in order that it be joined to the $t=const$ part (we have used a connector segment with approximately the same intrinsic geometry as on the first slice). The stretching is gentle and smooth, with all lengthscales and timescales associated to the stretching being of order $\sim M$. This stretching thus creates pairs of quanta on the slice, with wavelengths of order $\sim M$. Decomposing this state into quanta that emerge from the hole (labelled $b$) and quanta that stay inside (labelled $c$), we get a state that is schematically of the form $Exp[\gamma b^\dagger c^\dagger]|0\rangle_b\times |0\rangle_c$ for each step of the evolution. The explicit form of the state can be found for example in \cite{giddings} for the 2-d black hole. For our purposes we only need to note that this state is entangled between the inner and outer members of the pair, and so for simplicity we break up the evolution into a set of discrete steps (with time lapse $\Delta t \sim M$), and take a simple form of the entangled state having just two terms
\be
|\Psi\rangle_{\rm pair}=\sq \z_c\z_b+\sq\o_c\o_b
\label{pairs}
\ee
(Nothing in the argument below should be affected by this simplification, or the simplification of taking discrete timesteps.
If we had a fermionic field we would in fact have just two terms in the sum.) 

At the initial timestep we have on our spacelike slice only the shell that has passed through its horizon radius, denoted by a state $|\psi\rangle_M$. At the next time step we have, in the leading order Hawking computation, the state
\be
|\Psi\rangle= |\psi\rangle_M\otimes\Big( \sq \z_{c_1}\z_{b_1}+\sq\o_{c_1}\o_{b_1}\Big)
\label{qtwoq}
\ee
If we compute the entanglement of $b_1$ with $\{M, c_1\}$ we obtain 
\be
S_{ent}=\ln 2
\ee
At the next step of evolution the slice stretches so that the quanta $|b\rangle_1, |c\rangle_1$ move away from the middle of the `connector region' ${\cal C}$, and a new pair is pulled out of the vacuum near the center of ${\cal C}$.  The full state is
\bea
|\Psi\rangle=|\psi\rangle_M&\otimes&\Big( \sq \z_{c_1}\z_{b_1}+\sq\o_{c_1}\o_{b_1}\Big)\cr
&\otimes&\Big( \sq \z_{c_2}\z_{b_2}+\sq\o_{c_2}\o_{b_2}\Big)
\label{qtwoq2}
\eea
If we compute the entanglement of the set $\{b_1, b_2\}$ with $\{M, c_1, c_2\}$, we find
\be
S_{ent}=2\ln 2
\label{ent2}
\ee
Continuing this process, after $N$ steps we get, in the leading order Hawking computation,
\bea
|\Psi\rangle= |\psi\rangle_M&\otimes&\Big( \sq \z_{c_1}\z_{b_1}+\sq\o_{c_1}\o_{b_1}\Big)\cr
&\otimes&\Big( \sq \z_{c_2}\z_{b_2}+\sq\o_{c_2}\o_{b_2}\Big)\cr
&\dots&\cr
&\otimes&\Big( \sq \z_{c_N}\z_{b_N}+\sq\o_{c_N}\o_{b_N}\Big)
\label{qtwoq3}
\eea
The entanglement entropy of the  $\{ b_i\}$ with $ \{ M,  c_i\}$ is
\be
S_{ent}=N\ln 2
\label{ent}
\ee
Since this entanglement keeps growing with $N$, we get the Hawking problem mentioned above.

\subsection{Allowing small corrections}

Let the state at time step $t_n$ be written as  $|\Psi_{M, c, b}(t_n)\rangle$. Here $\{M\}$ denotes the  matter shell that fell in to make the black hole, $\{c\}$ denotes the quanta that have been created at earlier steps in the evolution and $\{b\}$ denotes the set of all $b$ quanta that have been created in all earlier steps. The state $\Psi$ is entangled between the $\{M,c\}$ and $\{b\}$ parts; it is not a product state. We assume nothing about its detailed structure. 

We can choose a basis of orthonormal states $\psi_n$ for the $\{M,c\}$ quanta inside the hole, and an orthonormal basis $\chi_n$ for the quanta $\{b \}$ outside the hole, such that 
\be
|\Psi_{M, c, b}(t_n)\rangle=\sum_{m,n} C_{mn} \psi_m\chi_n
\ee
It is  convenient to make unitary transformations on the $\psi_m, \chi_n$  so that we get
\be
|\Psi_{M, c, b}(t_n)\rangle=\sum_{i} C_{i} \psi_i\chi_i
\label{stateone}
\ee
We compute the reduced density matrix describing the $b_i$ quanta outside the hole
\be
\rho_{ij}=|C_i|^2 \delta_{ij}
\ee
The entanglement entropy at time $t_n$ is then
\be
S_{ent}(t_n)=-\sum_i |C_i|^2 \ln |C_i|^2
\label{entone}
\ee
In the leading order  evolution we would have  at time step $t_{n+1}$:
\be
|\Psi_{M, c, b}(t_n)\rangle\r |\Psi_{M, c, b}(t_n)\rangle\otimes \Big [\sqi |0\rangle_{c_{n+1}}|0\rangle_{b_{n+1}}+\sqi |1\rangle_{c_{n+1}}|1\rangle_{b_{n+1}}\Big ]
\label{leading}
\ee
where the term in box brackets denotes the state of the newly created pair. 

Let us now write down the modifications  to this evolution from timestep $t_n$ to timestep $t_{n+1}$ that will encode  the small corrections we wish to allow:

\b

(A) For the quanta that have left the hole at earlier timesteps, we have no change
\be
\chi_i\r \chi_i
\label{chifixed}
\ee
This is the case even for the example of burning paper: quanta that have left the paper and been collected outside do not participate in the next step of the evolution of the paper.

\b

(B) In the leading order evolution there was no change to the state of $\{ M,c\}$ inside the hole, since the part of the spacelike slice carrying these quanta did not evolve forwards in time. We will now allow a completely general unitary evolution of this state to any other state formed by these quanta. 

\b

(C) In the leading order Hawking computation the state of the newly created pair was $|\Psi\rangle_{\rm pair}$ (eq. (\ref{pairs})). We now allow the state of this pair to lie in a 2-dimensional subspace, spanned by $|\Psi\rangle_{\rm pair}\equiv S^{(1)}$ and an orthogonal vector $S^{(2)}$:
\bea
S^{(1)}&=&\sqi  |0\rangle_{c_{n+1}}|0\rangle_{b_{n+1}}+\sqi  |1\rangle_{c_{n+1}}|1\rangle_{b_{n+1}}\cr
S^{(2)}&=&\sqi  |0\rangle_{c_{n+1}}|0\rangle_{b_{n+1}}-\sqi  |1\rangle_{c_{n+1}}|1\rangle_{b_{n+1}}\cr
&&
\label{set}
\eea

\b

Putting together (B) and (C), we see that the most general evolution allowed for the state in the hole is 
\be
\psi_i\r \psi_i^{(1)}S^{(1)}+\psi_i^{(2)}S^{(2)}
\ee
where $\psi_i^{(1)}, \psi_i^{(2)}$ are any two states of $\{M,c\}$ (the initial matter and the infalling members of Hawking pairs produced at earlier steps). As promised in (B),  these states be arbitrary, but note that unitarity of evolution gives
\be
|| \psi_i^{(1)}||^2+|| \psi_i^{(2)}||^2=1
\ee
We thus get the evolution
\be
|\Psi_{M, c, b}(t_{n+1})\rangle=\sum_{i} C_{i} [\psi_i^{(1)}S^{(1)}+\psi_i^{(2)}S^{(2)}]~ \chi_i
\label{statetwo}
\ee

We can write the state (\ref{statetwo}) as
\be
|\Psi_{M, c, b}(t_{n+1})\rangle=S^{(1)}\Big [ \sum_{i} C_{i} \psi_i^{(1)} \chi_i\Big ] + S^{(2)}\Big [ \sum_{i} C_{i} \psi_i^{(2)} \chi_i\Big ] \equiv S^{(1)} \Lambda^{(1)}+ S^{(2)} \Lambda^{(2)}
\label{state}
\ee
Here we have defined the states
\be
\Lambda^{(1)}= \sum_{i} C_{i} \psi_i^{(1)} \chi_i, ~~\Lambda^{(2)}= \sum_{i} C_{i} \psi_i^{(2)} \chi_i
\ee
Since $S^{(1)}, S^{(2)}$ are orthonormal, normalization of $|\Psi_{M, c, b}(t_{n+1})\rangle$ implies that
\be
||\Lambda^{(1)}||^2+||\Lambda^{(2)}||^2=1
\ee

We can now state precisely what it means for the quantum corrections to be small. If the evolution from step $n$ to step $n+1$ is to be close to the semiclassical one in any sense then we must have produced mostly the state $|\Psi\rangle_{\rm pair}\equiv S^{(1)}$ and only a small amount if $S^{(2)}$. Thus we write
\be
|| \Lambda^{(2)}||<\epsilon, ~~~\epsilon\ll 1
\label{cond1}
\ee
If there is no such bound, then we will say that the corrections to the Hawking evolution are `order unity'. 

The result of \cite{mathurfuzz} says that with the smallness condition (\ref{cond1}) we get a minimal increase of entanglement at each step
\be
S_{n+1}-S_n>\ln 2-2\epsilon
\label{result}
\ee

It is important to note that the smallness condition (\ref{cond1}) is placed on the evolution inside a spacetime region that is size $\sim M$ on each side. Thus we are just writing down in mathematical terms the requirement that the full state and the semiclassical state differ by a small amount in the local process of evolution near the horizon by one timestep of order $M$ (the timescale required to create one Hawking pair). We are {\it not} requiring that the full state of all the $\{b\}$ quanta be close to the one that we would obtain by semiclassical evolution. For a state made of many quanta, or in the process of evolution by many timesteps, we can get large changes from the cumulative effects of small changes. For example suppose we have a product state $|\Psi\rangle=\prod_{i=1}^k \psi_i$, and another such state $|\Psi'\rangle=\prod_{i=1}^k\psi'_i$. Even if each $|\psi\rangle$ is close to the corresponding $|\psi'\rangle$ (i.e., $\langle\psi|\psi'\rangle=1-\epsilon$), we get a small overlap between the total wavefunctions
\be
\langle \Psi|\Psi'\rangle=(1-\epsilon)^k\r 0 ~~{\rm when}~~k\r \infty
\ee
By contrast, in our argument above we are only requiring that in a spacetime ball of size $\sim M$ the full evolution and the semiclassical one differ by a small amount, something that we {\it must} have if we are to say that the horizon evolution is close to the semiclassical one.  

\section{A simple model for correlations}
\label{model}\setcounter{equation}{0}

In this section we take a simple model for the corrections of order $\epsilon$ that we used above, and with these corrections, compute explicitly the entanglement entropy between the $\{b\}$ quanta outside the quanta inside the hole. We find, in accord with (\ref{result}), that the fractional reduction in entanglement is small, even though the number of emitted quanta $N$ is taken to infinity.

The essential structure of this model is as follows. At time step $n$ we have an amplitude to create a pair, and an amplitude that we do not create a pair. If we do create a pair, then the amplitude that we create a pair at the next timestep $n+1$ is enhanced a little (by a factor $e^a$, with $a\ll 1$). If we do not create a pair at step $n$, then the amplitude to create a pair at step $n+1$ is reduced a little (by a factor $e^b$, with $b<0, ~|b|\ll 1$).\footnote{The parameter $b$ should not be confused with the label $b$ used to denote the radiation quanta in the emitted pairs $\{c,b\}$.}Thus we will generate a correlation between the quanta emitted from the hole. Note that the enhancements and de-enhancements can only be slight, since the leading order Hawking process gives no correlations, and we are requiring that quantum gravity effects be small on the low energy modes involved in the Hawking process.

One way to get effects like the one described by this model is to include backreaction in the Hawking emission process. Each quantum in very light compared to the mass of the hole, so to leading order one usually computes the backreaction by putting in $\langle T_{\mu\nu}\rangle_{radiation}$ into the Einstein's equations to get an average backreaction from the emission process. Such a backreaction term has no effect on the entanglement problem, since it only results in a slow change in the background geometry, with entangled pairs being created as before at each step of the emission process. (For 1+1 dimensional models such a backreaction was explicitly computed in \cite{rst}, for example.) But we may wish to consider the backreaction in a more delicate way, computing the effect of each emission on the geometry instead of just using the  expectation value $\langle T_{\mu\nu}\rangle_{radiation}$ for the energy loss. In \cite{pw} the effect of radiation of a shell from the hole was computed. If at one step in the radiation process we have an emission of a shell, then the mass of the hole goes down a little and so the temperature goes up a little. This makes it slightly more likely that a shell will be emitted at the next time step. Conversely, if no shell was emitted on the earlier step, then we would find a slightly lower amplitude for the emission at the next step. This generates correlations among the emitted quanta, and it has been speculated that such correlations might be able to remove the information problem \cite{remove}. But we will see in our model below that the entanglement reduction due to these correlations goes to zero as $e^a\r 1, ~e^b\r 1$, so we cannot really solve the problem by small corrections.
The smallness of this reduction in entanglement is of course in accord with the general result (\ref{result}). 

\subsection{The model}

At the initial timestep ($n=0$) we have no created pairs; we can imagine that on this slice we have only the infalling shell that created the black hole. (We will not involve this matter in the further steps of the evolution of our simple model, but note that the general result (\ref{result}) holds with any arbitrary involvement of this matter.)

At the next time step ($n=1$) we create one pair of Hawking quanta, in the state (\ref{pairs})
\be
|\Psi\rangle_{n=1}=\sq \z_{c_1}\z_{b_1}+\sq\o_{c_1}\o_{b_1}
\label{pairsq}
\ee

At the next time step ($n=2$) we can again have no creation ($\z_{c_2}\z_{b_2}$) or the creation of a pair ($\o_{c_2}\o_{b_2}$). But the amplitudes for this process depend on whether a pair was created  at the previous step ($n=1$). If no pair was created at step $n=1$ then we have an amplitude $\sq e^a$ for no pair creation at $n=2$, and an amplitude $\sq e^b$ for pair creation:
\be
\z_{c_1}\z_{b_1}~\r~ \sq e^a\z_{c_2}\z_{b_2}+ \sq e^b \o_{c_2}\o_{b_2}
\ee
Note that by unitarity we have
\be
| \sq e^a|^2+| \sq e^b|^2=1
\ee
We will take $a, b$ real, so that 
\be
e^{2a}+e^{2b}=2
\label{sevent}
\ee
 We let $a>0$ (to allow enhancement of the emission amplitude if there was an emission at the previous step). From (\ref{sevent}) we see that $b<0$.

Smallness of the corrections implies that 
\be
|a|\ll 1, ~~~|b|\ll 1
\label{eightt}
\ee
In the computations that follow we will not use the approximation of small $|a|, |b|$, but at the end we will use the condition (\ref{eightt}) in arriving at our conclusion that small corrections give small changes to entanglement.

If we did have pair creation at step $n=1$, then we take amplitude to be $\sq e^b$ for no pair creation at the timestep $n=2$ and amplitude $\sq e^a$ for  pair creation:
\be
\o_{c_1}\o_{b_1}~\r~ \sq e^b\z_{c_2}\z_{b_2}+ \sq e^a \o_{c_2}\o_{b_2}
\ee
Thus the overall state at timestep $n=2$ is
\bea
|\Psi\rangle_{n=2}&=&\sq \z_{c_1}\z_{b_1}\otimes \Big [\sq e^a\z_{c_2}\z_{b_2}+ \sq e^b \o_{c_2}\o_{b_2}\Big ]\nn\cr
&+&\sq \o_{c_1}\o_{b_1}\otimes \Big [  \sq e^b\z_{c_2}\z_{b_2}+ \sq e^a \o_{c_2}\o_{b_2}
\Big ]
\eea
We continue in a similar way for each future time step. Thus we break up the state at timestep $n$ into two parts
\be
|\Psi\rangle_n=|\Psi\rangle_n^{(1)} \otimes \z_{c_n}\z_{b_n}+|\Psi\rangle_n^{(2)} \otimes \o_{c_n}\o_{b_n}
\ee
where the states $|\Psi\rangle_n^{(1)}, |\Psi\rangle_n^{(2)}$ involve the quanta $c_1, b_1, \dots c_{n-1}, b_{n-1}$. 
The state at timestep $n+1$ is then
\bea
|\Psi\rangle_{n+1}&=&|\Psi\rangle_n^{(1)} \otimes\Big [\sq e^a\z_{c_{n+1}}\z_{b_{n+1}}+ \sq e^b \o_{c_{n+1}}\o_{b_{n+1}}\Big ]\nn\cr
&+&|\Psi\rangle_n^{(2)} \otimes\Big [  \sq e^b\z_{c_{n+1}}\z_{b_{n+1}}+ \sq e^a \o_{c_{n+1}}\o_{b_{n+1}}
\Big ]
\eea
After $N$ timesteps  the state $|\Psi\rangle_N$ is an entangled state of the $\{ b\}$ quanta which constitute the radiation from the hole and the $\{ c \}$ quanta which are the members of the Hawking pairs that fell into the hole. We trace out the $\{ c \}$ quanta, getting a reduced density matrix $\rho^{(b)}$ describing the $\{ b \}$ quanta. We  compute the entanglement entropy
\be
S_{ent}=-Tr \rho^{(b)} \ln \rho^{(b)}
\ee
We then ask if $S_{ent}$ decreases after some point, or continues to increase with $N$. If it continues to increase with $N$ then we cannot solve Hawking's problem with such a model of small corrections.

\subsection{Description of the entangled state}

Consider the states of the $\{ b \}$ quanta. There are $N$ steps in the emission process, and at each step we either have no emission ($|0\rangle_{b_i}$) or an emitted quantum ($|1\rangle_{c_i}$). Thus we get a basis of orthonormal states of the $\{  b \}$ quanta by taking sets of $N$ numbers, with each number being a $0$ or a $1$. Thus there are $2^N$ possible states of the $\{ b \}$. We call these states $\psi_k$, labelled by an index $k=1, \dots 2^N$: 
We write
\bea
\psi_1&=&00\dots 00 \nn\cr
\psi_2&=&00\dots 01\nn\cr
&\dots&\nn\cr
\psi_{2^N}&=&11\dots 11
\label{tenq}
\eea

Now note that the radiation process produces particles in pairs, so whenever we have a $0$ in the $b$ space we also have a $0$ in the $c$ space, and when we have a $1$ in the $b$ space we have a  $1$ in the $c$ space. Thus the overall state of the $c,b$ system is of the form
\be
|\Psi\rangle_N=\sum_{k=1}^{2^N} C_k~ \psi^{(c)}_k\otimes \psi^{(b)}_k
\label{tw}
\ee
where the states $\psi_k$ are given in (\ref{tenq}), and the superscripts $c,b$ tell us which of the sets of quanta we are putting in that state. 

It is easy to write down the values of the $C_k$. Suppose $N=8$ and we have a state 
\be
\psi_k=11001010
\label{el}
\ee
The first entry on the left is $1$, which means that at the step $n=1$ we had emission of a quantum. At this initial step we see from (\ref{pairsq}) that we have amplitude $\sq$ for emission, so we get a factor $\sq$ in $C_k$. The next entry is again $1$, which would have an amplitude $\sq e^a$ since the emission of a quantum at the previous step enhances the amplitude for emission at the next step by $e^a$. The next entry is a $0$, and since this follows a $1$, we get a factor $\sq e^b$. Continuing in this way we get $N=8$ factors of $\sq$, and $N-1=7$ factors $e^a$ or $e^b$. For the above example (\ref{el})
\be
C_k={1\over 2^4}e^a e^be^a e^be^be^be^b
\ee
In general we  see that we get an overall  factor ${1\over 2^{N\over 2}}$, and then $N-1$ factors from the `links' between the numbers; we get $e^a$ for `no jump' and $e^b$ for a `jump'.

We now compute the density matrix for the $b_i$. From (\ref{tw}) we see that the density matrix is diagonal
\be
\rho^{(b)}_{kl}=|C_k|^2 \delta_{kl}
\ee
Finally, the entanglement entropy is given by 
\be
S_{ent}=-\sum_{k=1}^{2^N} |C_k|^2 \ln |C_k|^2
\label{thir}
\ee

\subsection{Mapping to the Ising model}

The fact that we have just two possibilities $0,1$ at each timestep suggests that we map the problem to a 1-d Ising model which has at each site a spin up ($\uparrow$) or a spin down ($\downarrow$). The state (\ref{el}) described an emission where at the first step we had occupation $1$ in the $c_1,b_1$ modes, at the next step we again had occupation $1$ in $c_2,b_2$ , then we had occupation $0$  in $c_3,b_3$ etc. We map such a state to a configuration of a 1-d Ising model with $N$ spins:
\be
11001010 \r \u\u\d\d\u\d\u\d
\label{fourt}
\ee
Thus when we sum over $k=1,\dots 2^N$ we will be going over all $2^N$ possibilities of spins in this Ising model.

Our goal is to compute the entanglement entropy (\ref{thir}). This involves the $|C_k|^2$, which have the following form. We have an overall factor ${1\over 2^N}$, and then a factor $e^{2a}$ when successive spins in a set like (\ref{fourt}) are the same, and a factor $e^{2b}$ when they are opposite. We write
\be
|C_k|^2= {1\over 2^N} e^{ w^{(k)}_1+ w^{(k)}_2+\dots w^{(k)}_{N-1}}
\ee
where $w^{(k)}_i=2a$ for `like spins' and $w^{(k)}_i=2b$ for `unlike spins' on any link.  

We have
\bea
S_{ent}&=&-\sum_{k=1}^{2^N} |C_k|^2 \ln |C_k|^2\nn\cr
&=&-\sum _{k=1}^{2^N} \Big[{1\over 2^N} e^{ w^{(k)}_1+ w^{(k)}_2+\dots w^{(k)}_{N-1}}\Big ] \Big [-N \ln 2 + (w^{(k)}_1+ w^{(k)}_2+\dots w^{(k)}_{N-1})\Big ]
\eea

Let us define
\be
Z[\mu]=\sum_{k=1}^{2^N} e^{\mu [ w^{(k)}_1+ w^{(k)}_2+\dots w^{(k)}_{N-1}]}
\label{fift}
\ee
Then we observe that we can write
\be
S_{ent}={N\ln 2\over 2^N} Z[1]-{1\over 2^N} \p_\mu Z[\mu]|_{\mu=1}
\label{sixt}
\ee

\subsection{Solving the model}

It is straightforward to solve for the partition function $Z[\mu]$ (eq. (\ref{fift})). The sum is over the $2^N$ spin configurations of a 1-d Ising model. The transfer matrix for this Ising model is
\be
T=\pmatrix {\d\r\d & \d\r\u\cr \u\r\d & \u\r\u}=\pmatrix {e^{2\mu a} & e^{2\mu b}\cr  e^{2\mu b} & e^{2\mu a}\cr }
\ee
where in the first step we have indicated what the matrix elements correspond to; for instance, the first entry in the matrix  corresponds to the case where  a down spin links to a down spin.

The eigenvalues and eigenvectors of this transfer matrix are
\bea
\lambda_1= e^{2\mu a}+e^{2\mu b}, &&~~~V_1={1\over \sqrt{2}}(1,1)\nn\cr
\lambda_2= e^{2\mu a}-e^{2\mu b}, &&~~~V_2={1\over \sqrt{2}}(1,-1)
\eea
By taking a copy of $T$ for each link and multiplying with the copy for the next link, we automatically perform the sum over spin states for the spins $2,3, \dots N-1$ (with the correct weights for the configurations). The sum over the end spins (numbered $1$ and $N$) can be included by multiplying the vector 
\be
V=(1,1)=\sqrt{2} V_1
\ee
at each end of the matrix product. Thus we have
\be
Z[\mu]=V_1^T T^{N-1} V_1= 2 V^T_1 T^{N-1} V_1 = 2 \lambda_1^{N-1}=2
(e^{2\mu a}+e^{2\mu b})^{N-1}
\ee

Putting this in (\ref{sixt}), and noting the relation  (\ref{sevent}), we get
\be
S_{ent}=N\ln 2-(N-1)[ae^{2a}+be^{2b}]
\ee
The entanglement in the leading order Hawking computation is obtained by setting $a=b=0$:
\be
S_{ent}^{leading}=N \ln 2
\ee
Thus the fractional correction in entanglement due to the correlations is
\be
{|\delta S_{ent}|\over S_{ent}}={(N-1)[ae^{2a}+be^{2b}]\over N\ln 2}
\ee
Recall that for our physical problem we wish to keep $|a|\ll1, ~|b|\ll1$ (eq. (\ref{eightt})). Then the constraint (\ref{sevent}) gives
\be
b\approx -a
\ee
and we find for $N$ large
\be
{|\delta S_{ent}|\over S_{ent}}\approx {2 a^2\over \ln 2} \ll 1
\ee

We see that the correction to the leading order Hawking computation is fractionally very small. Thus, in accordance with the general result (\ref{result}), we find that corrections to the Hawking process of the type we assumed cannot remove the entanglement between the radiation and the quanta in the hole.

\section{Consequences of the inequality (\ref{result})}
\label{cons}\setcounter{equation}{0}

The inequality (\ref{result}) establishes that the entanglement leading to Hawking' problem cannot be removed by including small corrections from quantum gravity or other sources. Thus the Hawking problem is indeed an important one, and has several consequences for the way we must think about gravity and the nature of black holes. We discuss some of these issues in this section.

\subsection{Separation of the `information paradox' and the `infall problem'}

A common way of posing Hawking's problem is the following: ``The horizon is a smooth place in the black hole geometry.  Thus an infalling observer sees nothing special at the horizon, and if there is nothing at the horizon then the outgoing radiation cannot encode any information.''
We will argue that this way of stating the issue confuses two different (and in a sense, opposite) problems.

The inequality (\ref{result}) was derived in \cite{mathurfuzz}, where it was placed in the context of the general information problem of black holes. Because of this inequality one finds the following: either we have {\it order unity} corrections at the horizon for the low energy modes contributing to Hawking radiation, or we must necessarily have information loss/remnants. The possibility of information loss and  the possibility of remnants cannot be separated in this argument, because the argument does not address what happens when the black hole becomes planck size. But the argument does exclude the fact that small quantum gravity effects can cause the information to leak away as subtle correlations in the Hawking radiation. Thus for quanta that have energy at infinity of order the Hawking temperature
\be
E_{rad}\sim kT
\label{qone}
\ee
we must find corrections of order unity from `nontrivial structure' at the horizon region, if we are to have any hope of avoiding information loss/remnants.

But this fact does not directly address the issue of what will be seen by an infalling observer. Imagine such an observer starting his infall at infinity. If his energy $E_{obs}$ at this point is order $kT$, then at all points of the infall his quanta can be confused with quanta in the radiation bath that he is immersed in, and there will be an order unity thermal randomness in any measurements that he tries to make. Thus we would like the observer (and his equipment) to be made of quanta with
\be
E_{obs}\gg kT
\label{qtwo}
\ee
We can now ask if quanta satisfying (\ref{qtwo}) might behave in a way where they see a horizon that is effectively a smooth place. We will discuss below  how this might happen, but for now we note that the problem of what happens at the horizon splits into two problems:

\b

(i) {\it The information paradox:}\quad To avoid information loss/remnants we must find some way of producing order {\it unity} corrections to the low energy modes at the horizon (these modes (\ref{qone}) have wavelength of order the black hole size in `good' coordinates at the horizon). This requirement is equivalent to finding `hair' at the horizon, so we are back to the difficulty of constructing solutions to the gravity theory that are not just the vacuum state at the horizon. If we do find such `hair', then this hair will encode the information of the black hole state, and transfer it to the radiation just like a piece of burning paper transfers the information in the paper's atoms to the outgoing photons. Thus the information paradox can be solved if (and only if) we can construct such `hair' for the black hole.

\b

(ii) {\it The infall problem:}\quad  Infalling observers have high energy (\ref{qtwo}). We can ask for the behavior of such high energy quanta  when they fall onto the degrees of freedom that describe the `hair'. It is possible that these quanta excite collective modes of the complicated hair states which can be described in an effective way by smooth infall through a horizon. We call this the `infall problem'. Note that there are two time scales in the black hole problem: the short `crossing timescale' $t_c$ and the very long `Hawking evaporation timescale' $t_H$. The infall problem only asks for the evolution of these high energy quanta on the timescale $t_c$; over the long timescale $t_H$ we will need the detailed structure of the hair to compute the evolution of the excitation and to understand how its information gets transferred to the outgoing radiation. 

\b

In this sense the `information paradox' and the `infall problem' are  `opposite problems'. To solve the information paradox we must construct `hair' which will modify the radiation process to an extent that each state of the black hole will lead to a {\it different}, unentangled state of the radiation. To solve the `infall problem' we must find a coarse grained description of high energy excitations which shows how different  hair states behave in a way that is effectively the {\it same} under these high energy, short time processes. 

For an analogy, consider a room full of air. 
If we wish to consider the motion of a single molecule in this room, we will have to know the positions of all other molecules, and compute the Brownian motion of our molecule as it scatters off the other molecules.  Now consider the motion of an apple thrown across the room. For such a heavy object (the mass of the apple is much more than the mass of individual air molecules) we can get a good approximation to the motion by replacing the air by a fluid with a small viscosity, or perhaps by ignoring the air altogether. We may also wish to see what happens over a time much longer than the crossing time of the apple across the room (this longer time is analogous to the long time $t_H$). The apple eventually decays, and how the resulting molecules get transported out through cracks in the room windows depends sensitively on the presence of air in the room and its precise state.  

\subsection{AdS/CFT duality and the information paradox}

One of the deep insights from string theory is the AdS/CFT correspondence \cite{maldacena,gkp,witten}.
It is sometimes argued that we can resolve the information paradox just by using the idea of AdS/CFT duality. It was shown in \cite{mathurfuzz} that such an argument is in fact not possible, due to the inequality (\ref{result}). Let us review that argument here; below we will note that the other question -- the infall problem -- {\it can} be related to the central idea of AdS/CFT.

Quantum mechanics is observed to be true in numerous experiments in physics, chemistry and even astronomy. Hawking argued for a breakdown of unitarity only when a black hole forms and evaporates. In this evaporation problem he gives an explicit computation showing the breakdown of unitarity, and the inequality (\ref{result}) shows that no small correction can invalidate his argument. Thus one cannot point to the {\it other} places (i.e., non-black hole situations) and argue that since quantum theory works there, it must also work in the black hole.

In the case of the AdS/CFT correspondence the situation is similar. There are a large number of computations where we can check the correspondence but where we do not make black holes. For example we can check operator dimensions, 2,3 and 4 point functions, Wilson loops etc., and observe a match between the gravity theory and the CFT. But noting that quantum theory works well for these gravity processes does not allow us to conclude that if we make a black hole in the same theory, then the correspondence will continue to hold, because this is the point at which Hawking's argument will appear to obstruct the relation between a gravity theory black hole and a manifestly unitary CFT. More precisely, suppose we describe the black hole in AdS by writing down the AdS-Schwarzschild metric for $AdS_5\times S^5$:
\be
ds^2=-(r^2+1-{C\over r^2})dt^2+{dr^2\over r^2+1-{C\over r^2}}+r^2 d\Omega_3^2 \times d\Omega_5^2
\label{qfour}
\ee
then we can draw `good slices' in the geometry just like we did for the metric (\ref{qthree}), and find again an inequality like (\ref{result}) showing that we will be forced to information loss/remnants even when arbitrary small corrections to the evolution are included. Thus we must either accept information loss (which leads to a breakdown of the map between gravity and the manifestly unitary gauge theory), or accept remnants in the gravity theory, or find some way to produce order unity corrections to the low energy outgoing Hawking modes by constructing the `hair' which modifies (\ref{qfour}). Thus if we do not want information loss/remnants, then we are back to the problem of trying to construct hair, something which has proved difficult in the past.

One may take a slightly different approach, and simply define the gravity theory as the dual description of the unitary CFT. Whatever this gravity theory gives, it certainly cannot have information loss since it is unitary. But let us look at our options again:

\b

(i) This gravity theory will behave like an expected theory of gravity in the computation of low energy processes (like the $n$-point functions for small $n$), but there may be no formation of a long lived black hole when we go to high energies. An example of where such a situation occurs is the 1-dimensional matrix model, which maps to a 1+1 dimensional string theory \cite{dasjevicki}.  This matrix model can be reduced to its singlet sector, which can in turn be described by fermions filling a fermi sea. Collective excitations of this fermi sea can be mapped to excitations in a 1+1 dilaton gravity theory, where one finds the expected `gravitational interaction time delays' for low energy processes \cite{folds1}. But if we make a collective excitation that might be strong enough to make a black hole, then we get `folds' in the fermi sea (which can be understood in terms of large interactions between the large number of quanta in the excitation), and we do not end up making any long lived black hole type object \cite{folds1,folds,smmm}.

\b

(ii) We may get remnants, which either slowly leak out their energy and information to infinity or survive as stable states for all time. Such a situation would obtain for example in the CGHS model \cite{cghs}, if we imagine that non-unitarity is not an option. The extended Penrose diagrams in loop quantum gravity also show information leaking out after the nominal endpoint of evaporation; information does not come out in the Hawking radiation \cite{loops}. More generally, this would be the situation in any gravity theory  
where we do not have information loss but where we also cannot find hair that would modify the evolution of low energy modes by order unity.

\b

(iii) We get the information encoded in the outgoing Hawking quanta. Weak coupling computations suggest that Hawking radiation is a unitarity process \cite{radiationall}, but at the couplings where we get a good gravity black hole  the Hawking argument for information loss (together with the result (\ref{result})) presents an obstacle. To overcome this obstacle we have to show how order unity corrections can arise to the Hawking process; in other words we are back to the issue of having to construct `hair' for the black hole.

\b

To summarize, just invoking the idea of AdS/CFT duality cannot resolve the information problem. To understand the information problem we have to not only know if unitarity holds in gravity, but in case it holds, to also distinguish between the three options (i), (ii), (iii) above. 

In this context we see the significance of the inequality (\ref{result}).  If small corrections to the radiation process could remove the entanglement and transfer information then we could content ourselves with just writing the metric (\ref{qfour}) for black holes in AdS, and assume that small quantum gravity effects (which would always be present) would take care of the information issue. If such was the case, then there would be no information paradox in the first place and nothing to worry about: why bother to see exactly how the information emerges as long as it is clear that there exist effects that can make it emerge? The power of the information paradox is that small corrections {\it cannot} get the information out, and resolving the paradox would change in a basic way our understanding of how quantum gravity effects can operate across macroscopic distances.

\section{The fuzzball construction, the information paradox and the infall problem}
\label{fuzzball}\setcounter{equation}{0}

In string theory one finds that there is indeed a complete set of `hair' for black holes. In this section we briefly review the construction of this hair, discuss how it resolves the information paradox, and conjecture how it relates to the infall problem.

\subsection{Black hole microstates in string theory (Fuzzballs)}

In string theory it turns out that we {\it do} find hair for the black hole; this hair gives order unity corrections to the evolution of modes at the surface of the black hole microstate, thus avoiding the inequality (\ref{result}). Let us review the nature of this hair.

String theory lives in 10 spacetime dimensions, and if we make a black hole in 4+1 dimensions, say, then there will be 5 compact dimensions \cite{sv}. Of course the idea of compact extra dimensions is an old one, but it was not obvious how such dimensions could provide hair. To get the traditional solution for the black hole one assumes spherical symmetry in the angular directions and also no dependence on the compact coordinates $y_i$. One then gets a solution with a horizon, and a singularity inside this horizon. Small perturbations to this solution do not give any regular solutions; this is the origin of the `no hair' conjecture \cite{nohair}. Compact directions do not seem to help; after all small perturbations from these directions decompose into scalars, gauge fields etc, and all such fields were explored in the search for hair.

How then do we end up with hair in string theory? It turns out that the hair arises from {\it nonperturbative} effects involving the compact directions. Consider a single compact circle. We can make a nontrivial but regular topological construct - the KK monopole \cite{kk} -- by fibering this circle over the noncompact directions. The microstates for the simplest hole in string theory (The D1D5 extremal hole)  arise from the existence of a structure that is topologically described by a KK monopole times an $S^1$  path in the noncompact directions \cite{lm4,2charge}. There is no net KK charge because the KK charge from one side of the $S^1$ path is cancelled by the charge from the `antimonopole' at the other side of this circular path. Different shapes of this $S^1$ give different microstates. 

\begin{figure}[htbp]
\begin{center}
\includegraphics[scale=.08]{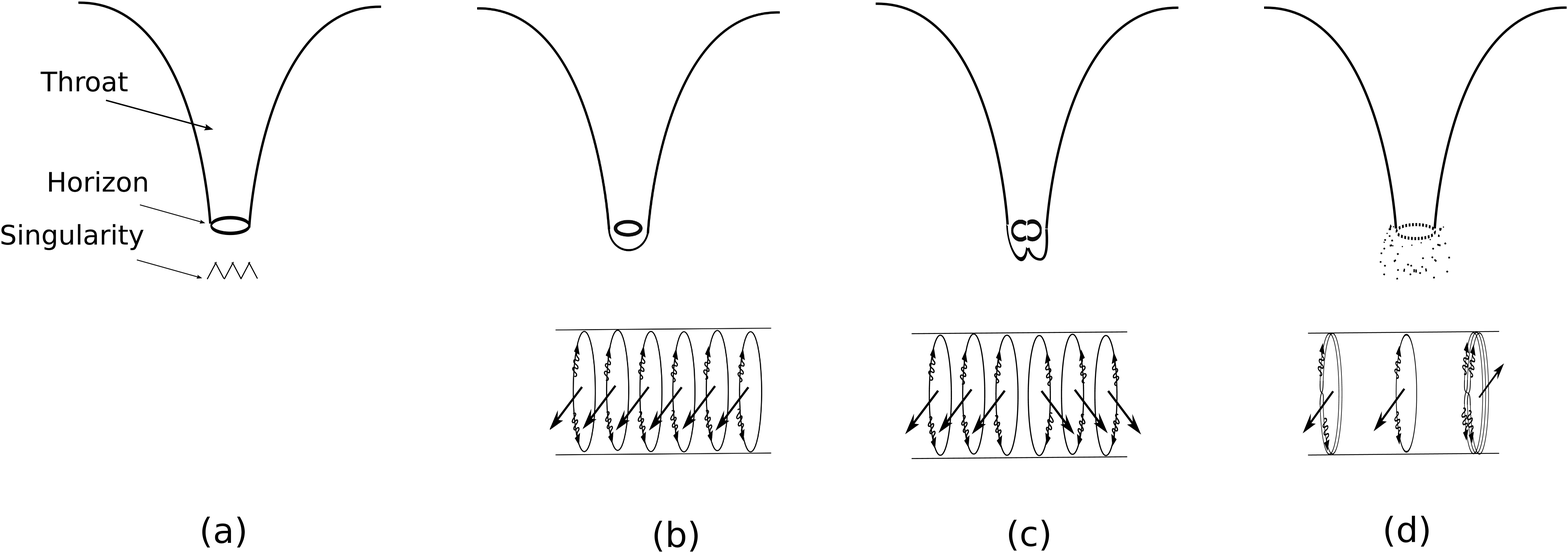}
\caption{{(a) The traditional black hole geometry: the state at the horizon is the vacuum (b) The simplest microstate (which has the $S^1$ in the shape of an exact circle), and the CFT state that it corresponds to; all `bits' are in the same state (c) The next simplest geometry; the CFT state has its bits distributed over two different types (d) A generic state, very quantum, but with no traditional horizon; the CFT state has bits distributed over many different types.}}
\label{second}
\end{center}
\end{figure}

To understand better the set of all states we can look at the list of states in the dual CFT. We can order these states in some rough order of `complexity'. The simplest states have all `bits' in the same state (this is depicted in fig.\ref{second}(b); for a detailed understanding of what these bits are, see \cite{mathurreview}). The corresponding gravity solution has no horizon and no singularity, and has very low quantum fluctuations. The $S^1$ path in the noncompact directions  is an exact circle. We can then move on to the next more complicated CFT state (fig.\ref{second}(c)), where the `bits' in the CFT are split among two possible states; the gravity solution is again regular but slightly more `complicated', in the sense that the $S^1$ describes a curve that is more `wiggly' than an exact circle.

For an analogy of how we are listing states, consider  the electromagnetic field in a box. The simplest state has all photons in a single mode, giving a `laser beam'. The quantum fluctuations are small, and the solution close to classical. The next simplest state has its excitations split between two modes, and so on. Proceeding this way, we finally end up with generic states where the occupation number per mode is order unity, and quantum fluctuations are order unity as well (fig.\ref{second}(d)). 

A similar situation holds in the gravity case. As we move to more complex microstates, we have less `bits' of each type (and therefore more types of bits), until we reach generic states where we have order unity bits of any given type. The  $S^1$ in the gravity solution becomes extremely `wiggly', to the point where the solution cannot be reliably described by any  metric. We will discuss quantum effects in such states below, but for now we wish to focus on the main point: as we move down the family of microstates (starting from the simple ones towards the more complicated ones) we do not find at any stage the traditional geometry with a regular horizon (i.e. a horizon where the state of quantum fields is the vacuum). 
That is, in each microstate that we construct the evolution of modes at the surface is different by {\it order unity} from the evolution of modes at a traditional horizon. Alternatively, we can try to write a state describing the fields at the location where a horizon would have appeared in the naive geometry, and then we find that this state $|\psi_k\rangle$ is close to {\it orthogonal} to the vacuum 
\be
\langle 0|\psi_k\rangle\r 0 ~~ {\rm as} ~~{M\over m_{pl}}\r\infty
\ee
 instead of the situation with the traditional horizon where the state $|0\rangle_H$ at the horizon is almost {\it equal} to the vacuum
\be
|\langle 0 | 0\rangle_H-1|\r 0~~{\rm as }~~ as ~~ {M\over m_{pl}}\r\infty
\ee

To summarize, string theory provides us a new expansion parameter: the `complexity' of the state.  Moving from simpler states towards more complicated states does {\it not} bring the state at the boundary of the `fuzzball' closer to the vacuum state that we get at the horizon of the traditional hole; we remain order unity away from the vacuum state. Extrapolating to the generic state brings in large quantum corrections, but we do not find any approach to the traditional black hole horizon. Thus we get `hair' for the black hole.

  \begin{figure}[htbp]
\begin{center}
\includegraphics[scale=.15]{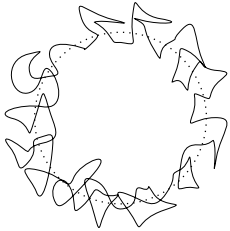}
\caption{Schematic description of a microstate solution of Einstein's equations. There are `local ergoregions' with rapidly changing direction of frame dragging near the horizon. The geometry closes off without having an interior horizon or singularity due to its peculiar topological structure. Generic microsates are very quantum, and so this figure just gives a schematic description of the rapid fluctuations near the fuzzball surface.}
\label{ftwop}
\end{center}
\end{figure}

\subsection{Fuzzballs and the information paradox}

From what we noted above, the states of string theory do not generate the horizon (with its local vacuum) that Hawking used in his argument. Thus we evade the trap of the information paradox.

To understand the Hawking radiation process with these microstates, we can return to the simplest microstates and ask if they radiate energy. It turns out that the nonextremal states have an ergoregion \cite{ross}, and the ergoregion emission \cite{myers} {\it exactly} equals the radiation we expect from the corresponding microstate based on the Hawking radiation computation in the dual CFT \cite{cm1}. For a discussion on how this process is expected to look for more generic states, see \cite{cm2}.

Finally, one may ask what happens when a shell collapses through its horizon to make a black hole.
A crude estimate indicates that the wavefunction of the shell simply spreads over the very large phase space of fuzzball states. One can think of this process as one of tunneling from the shell state to the fuzzball states: the amplitude for such a tunneling is very small ($\sim Exp[-GM^2]$), but the number of states that we can tunnel to is very large
($\sim Exp[S_{bek}]\sim Exp[GM^2]$) \cite{rate}, and so compensates for the smallness of the tunneling amplitude.  Another simple estimate shows that this tunneling can happen in a time much shorter than the Hawking evaporation time, so we find that the collapsing shell wavefunction evolves to a linear combination of fuzzball states and then radiates like a normal warm body from its surface \cite{time}. 

Putting together these results, we get a picture for resolving the information paradox. The infalling matter should be decomposed into eigenstates of the full Hamiltonian of the theory. These eigenstates are `fuzzballs'; i.e., eigenstates of the gravity theory do not possess a traditional horizon. The wavefunction of a collapsing shell spreads over the space of fuzzball states, which then leak  information by radiation in just the same way that  information emerges in the radiation from the surface of a piece of burning paper. 

\subsection{The infall problem}

We have seen above how microstates in string theory differ from each other at the horizon. The fuzzball structure is fluctuating on a very fine scale for generic microstates (fig.\ref{ftwop}). We can hope that the infall of a high energy quantum $E\gg kT$ (eq. (\ref{qtwo})) can excite collective modes of this fuzzball, giving a behavior that is essentially  the {\it same} for all generic fuzzballs. Such collective behavior of a large number of degrees of freedom is encoded in the basic idea of the AdS/CFT correspondence, and in its precursor, the matrix models.

\subsubsection{The `Dipole charge' structure of fuzzballs}

For the 2-charge D1D5 system discussed above, one sees that a KK monopole charge appears locally, but the net KK charge is zero. Such charges are called `dipole charges' and their appearance prevents the solution from acquiring a horizon and singularity. But the KK monopole is only one of a set of `branes' in string theory (under a `9-11 flip' the KK monopole maps to a D6 brane), and more generally all the branes of the theory can appear as dipole charges. 3-charge and 4-charge fuzzball solutions have been explicitly constructed, and different microstates differ in the arrangement of dipole charges \cite{3charge}. 

\subsubsection{Excitations of the fuzzball}

We are concerned now with the excitations of these fuzzballs. There are several kinds of excitations that one may list:

\b

(i) Gravitons, and other string modes living in the regions between the dipole charges

(ii) Fluctuations in the dipole charge locations and orientations

(iii) Creation of new sets of dipole charges

(iv) Excitations `linking' dipole charges. For an example, consider the D1D5 case with compactification $M_{4,1}\times T^4\times S^1$. In the KK monopole times $S^1$ dipole structure of the microstates, we can wrap a D3 branes as follows: one direction on the compact $T^4$, and two on the $S^2$ that links one the KK monopole at one point of the $S^1$ to the KK monopole at another point of this $S^1$.

\b

\begin{figure}[htbp]
\begin{center}
\includegraphics[scale=.15]{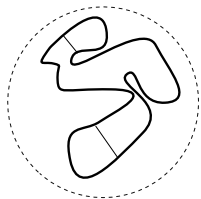}
\caption{The KK monopole times $S^1$ dipole structure of the 2-charge D1D5 fuzzball. Line segments joining points on this $S^1$ correspond to branes wrapping the sphere between KK monopoles.}
\label{ffive}
\end{center}
\end{figure}

More generally, we will not be concerned with the details of the excitations, but with the fact that there are a large number of closely spaced excitations, which are such that they can be excited by an infalling quantum with $E\gg kT$. For excitations of type (iv) above we can dualize the KK monopole `tube' to a D6 brane, whereupon the branes stretched between its points become open strings linking different points of the convoluted D6 brane. The mass of these open strings depends on the distance between the points that are linked, so low energy quanta cannot excite them but the high energy quanta see them all as essentially massless string modes. In fact the different segments of the D6 brane in this case form a 1/4 BPS set of `branes at angles', and the open strings between these branes gives the general excitation of such a 2-charge system.

Thus we see that the complicated structure of the fuzzball leads to a large set of closely spaced excitations of the {\it dipole} degrees of freedom. We can develop a membrane paradigm \cite{membranep} with these real degrees of freedom \cite{membrane}, instead of imaginary ones that we would have to conjecture if we had the traditional horizon with the vacuum state of matter. Collective modes of these degrees (translation, shear, addition of energy that locally moves the fuzzball boundary outwards, etc.) can be studied as excitations of these fuzzball states. 

\subsubsection{Thermalization in the dual CFT}

At this point it is useful to note how an infalling quantum evolves in the dual CFT description. Suppose a scalar quantum $h_{12}$ falls onto a D1D5 bound state, where $h_{12}$ is a graviton with indices in two of the compact $T^4$ directions. The CFT state is of the kind depicted in fig.\ref{second}, and one of the strands of the effective string picks up a right moving and a left moving excitation. It is important that when the incoming $h_{12}$ quantum interacts with these strands, all strands have (to a first approximation) the {\it same} amplitude to get excited. We say that such an excitation lies in the `symmetric sector' (S) of the CFT. Note that the state in fig.\ref{second}(b) has {\it only} a symmetric excitation at this order; since all strands are identical it makes no sense to say that we have different amplitudes for exciting different strands. For the CFT state in fig.\ref{second}(c) we can excite both types of strands with the same amplitude (S sector state) or have a positive amplitude for exciting one set and a negative amplitude for the other set. 
The latter excitation would  lie in the `antisymmetric sector' (A). Such an excitation has the same energy, but describes modes living deep in the fuzzball `cap' region of the dual geometry. 

The initial excitation is in the S sector, but interactions in the CFT (generated by the twist interaction operator) move it progressively to the A sector \cite{averypapers}. As noted in \cite{lm4}, the formation of a `black hole' is characterized in the CFT by all strands of the effective string being occupied by excitations. Approaching a black hole `horizon' in the gravity theory is described in the CFT by the excitations spreading to energies that are typical of the generic excitation on the effective string. While infall in empty AdS is described by a gradual move from the S sector towards A sector degrees of freedom,  reaching near a black hole horizon is characterized by a rapid transition to the A sector due to the fact that all strands are occupied by random excitations at this energy scale, and the wavefunction on the initial quantum is thus pulled into different waveforms on different strands (thermalization).

\subsubsection{The dual description of collective excitations}

A quantum falling towards the `horizon' of the black hole\footnote{This `horizon' is just the surface of the fuzzball \cite{lm5}.} will be compressed in the radial direction (due to the large redshift near the `horizon') but still be a `flat' wavefunction in the directions along the surface of the horizon. Thus it interacts with a large number of the closely spaced degrees of freedom at the surface of the fuzzball, a situation which leads to its absorption into the closely spaced fuzzball degrees of freedom by the `fermi golden rule'. We can now ask for an effective description of this absorption into the collective modes of the fuzzball. In the 1-dimensional matrix model \cite{dasjevicki}, one finds a smooth evolution for the collective modes of a large number of interacting degrees of freedom. This idea also appears in Matrix theory \cite{bfss}, where the effective dynamics of a large number of interacting degrees of physics encode all the dynamics of string theory. Of course the most clear manifestation of this collective dynamics arises in the AdS/CFT correspondence, where the degrees of freedom of a large number of branes gives rise to effectively smooth dynamical infall into an AdS type geometry \cite{maldacena}. 

Finally, we can come to the role of the traditional black hole metric. This metric does not arise as any of the solutions of string theory for given mass and charge.  But it can arise in the following effective ways:

\b

(i) We can rotate to Euclidean time, compactify this time and thus compute a thermal partition function of all fuzzball states. This path integral may have a saddle point, and even though the fuzzballs were individually not spherically symmetric, the saddle point is expected to be just the spherically symmetric Euclidean black hole geometry \cite{mathurreview}.

\b

(ii) We have seen that infall of an energetic quantum onto the generic fuzzball leads to the excitation of a large number of degrees of freedom of the fuzzball, so that we have `thermal physics' at this membrane. For a sufficiently complicated system we can replace the microstate with its ensemble average, which can be incorporated in the `real time formalism'  of thermal field theory by a second copy of the system and the choice of a particular entangled state of this pair of systems \cite{umezawa}. In the black hole case we can understand the thermodynamics of the hole by letting the two halves of the eternal black hole diagram be the two entangled systems of thermo-field-dynamics \cite{israel2}. In the dual CFT, the entangled state of two copies of the CFT should correspond to this eternal black hole geometry \cite{maldacena2}. It has recently been argued that summing over many geometries in two different spacetimes is effectively equivalent to a geometrical connection between the otherwise disconnected spacetimes \cite{raamsdonk}.

Thus we get the following possible picture for describing the collective dynamics of fuzzballs. A heavy infalling object excites many degrees of freedom of this fuzzball, which can be studied in a `thermal' way. For such a situation we can replace the fuzzball by an ensemble average. To do this, we first take the set of all fuzzballs, another copy of this set, and entangle the two sets. This would be equivalent to a geometry where we can now smoothly extend to the other side of an effective horizon. Such a picture would manifest the idea of `complementarity' \cite{complementarity}, where the interior of the hole arises from an effective description of the degrees of freedom that are seen by an external observer.

Thus once we have real degrees of freedom at the horizon, we can study their coarse grained description  relevant to the infall of a heavy  `observer'. But the information paradox needs us to first find these degrees of freedom, since small quantum corrections do not provide the effects necessary to give a unitary theory. 

\section{Discussion}

We have presented an example of small corrections to the Hawking emission from a black hole (roughly modelling the suggestion of \cite{pw}). Since the model was a simple one, we could solve it in closed form and compute the entanglement between the outgoing radiation quanta and their partners inside the hole. In accordance with  the general result of \cite{mathurfuzz}, these small corrections fail to have a significant impact on the entanglement which gives rise to Hawking's information puzzle.

One may also wish to look at explicit models where the outgoing quanta get affected by more states than just the one at the previous step of emission (the general inequality of \cite{mathurfuzz} allows for the most general such interaction). Since is it hard to solve more complicated models analytically, one can explore the problem numerically by evolving the system with any chosen rule for correlations. The results from these computations follow a similar pattern, and will be presented elsewhere.

We then discussed the importance of the inequality (\ref{result}) for the information problem. If small corrections could remove the entanglement between inner and outer members of the Hawking pairs, then there would be no information paradox in the first place; one would just take the metric (\ref{qthree}) as a leading order description of the hole, and let the quantum mechanics be a self-consistent problem of small fluctuations around this background. In the AdS context, we would similarly write the metric (\ref{qfour}), assume that small stringy effects ensure unitarity, and proceed to compute effects like particle infall by using a coarse grained description in the dual CFT (e.g. by analyzing the dual CFT in the hydrodynamic limit \cite{minwalla}). But small corrections   do {\it not} lead to unitarity, forcing us back to the search for order unity corrections at the horizon (`hair'). It turns out that string theory {\it does} have such hair; this hair arises from nonperturbative effects involving the extra dimensions and the branes of the theory, and so would not be found in a theory like canonically quantized 3+1 dimensional gravity or the CGHS model. Thus we expect the latter theories to give  remnants or information loss, while string theory can have the information leak out in the Hawking radiation. 

In short, the inequality (\ref{result}) separates the issues with black holes into two problems. The information paradox asks for how entanglement caused by Hawking radiation be removed from the black hole, and to answer this we must explicitly find the structures in our gravity theory that will give order unity corrections to the behavior of low energy modes at the surface of the hole. In string theory we find the `fuzzball' structure which achieves this task. The other problem is then the `infall problem' which asks for the coarse grained behavior of the degrees of freedom at the horizon and if these can be represented in some way by smooth effective degrees of freedom.\footnote{See \cite{bala} for discussions on some ways to get effective geometries from microstates.} This latter problem can likely be addressed by the collective field ideas embodied in matrix theories and AdS/CFT.

\section*{Acknowledgements}

I am grateful to Sumit Das, Steve Giddings, Stefano Giusto, Patrick Hayden, Gary Horowitz, Werner Israel, Don Marolf, Don Page, Christopher Plumberg, Ashoke Sen and Edward Witten for discussions on various aspects of this problem.  This  was supported in part by DOE grant DE-FG02-91ER-40690.

\end{document}